# Binary Atomic Silicon Logic

Taleana Huff[1,3,*], Hatem Labidi[1,2], Mohammad Rashidi[1], Lucian Livadaru[3], Thomas Dienel[1], Roshan Achal[1,3], Wyatt Vine[1], Jason Pitters[2,3], and Robert A. Wolkow[1,2,3*]

[1]Department of Physics, University of Alberta, Edmonton, Alberta, T6G 2J1, Canada
[2] Nanotechnology Research Centre, National Research Council Canada, Edmonton, Alberta, T6G 2M9, Canada
[3]Quantum Silicon, Inc., Edmonton, Alberta, T6G 2M9, Canada
*Correspondence to: taleana@ualberta.ca rwolkow@ualberta.ca

**It has long been anticipated that the ultimate in miniature circuitry will be crafted of single atoms. Despite many advances made in scanned probe microscopy studies of molecules and atoms on surfaces, challenges with patterning and limited thermal stability have remained. Here we make progress toward those challenges and demonstrate rudimentary circuit elements through the patterning of dangling bonds on a hydrogen-terminated silicon surface. Dangling bonds sequester electrons both spatially and energetically in the bulk band gap, circumventing short-circuiting by the substrate. We deploy paired dangling bonds occupied by one moveable electron to form a binary electronic building block. Inspired by earlier quantum dot-based approaches, binary information is encoded in the electron position allowing demonstration of a "binary wire" and an OR gate.**

The prospect of atom scale computing was initially indicated by "molecular cascades" where sequentially toppling molecules were arranged in precise configurations to achieve binary logic functions [1]. Many notable approaches toward molecular electronics [2–7], atomic electronics [8,9], and quantum-dot-based electronics [10–15] have also been explored.

The quantum dot based approaches [16–20] are particularly attractive, as they provide a low power yet fast basis [21] to go beyond today's CMOS technology [22]. These approaches, however, require cryogenic temperatures to minimize the population of thermally-excited states and achieve the desired functionality. Variability among quantum dots and sensitivity to uncontrolled fields are known to pose additional challenges [23]. The prospect to partially circumvent these issues was reported in studies of silicon dangling bonds (DBs), *i.e.*, unsatisfied bonds, on the otherwise hydrogen-terminated silicon surface (H-Si) [15].

Silicon DBs behave like quantum dots because they are zero-dimensional and exhibit three distinct charge states (positive, neutral, and negative) depending on their electron occupation (zero, one, or two electrons, respectively) [24,25]. Consequently, DBs have two charge transition levels, namely the neutral to negative (0/-) and positive to neutral (+/0) transitions. Crucially, because these DB energy levels lie within the bulk bandgap they are electronically isolated from the bulk [25,26]. Silicon DBs approach the ultimate small size (single atom) for a quantum dot and therefore exhibit larger energy level spacing, relaxing temperature requirements compared to larger conventional quantum dots [13]. It has additionally been established that H atoms and DBs on the silicon surface are stable against diffusion even at 200 °C, corresponding to a diffusion barrier of 1.4 eV [15,27,28]. DBs can be patterned at precise lattice locations allowing their positions and spacing's to be exactingly defined [29]. Because all dots are identically composed of only one atom, inhomogeneities are limited to local environment variability, which in principle can be effectively eliminated.

H-Si was first identified as an attractive candidate for nanoscale patterning by Lyding *et al.* [30]. Only recently have capabilities reached the levels necessary to enable prototyping of structures on this surface. DBs can now be deterministically placed or erased (controlled H atom placement) using a scanned probe [29,31,32]. Recently, prospects for atom scale fabrication have improved through the application of machine learning methods to automate some of the most challenging aspects of scanned probe atom-scale imaging and fabrication [33].

Using these new methods, we demonstrate that two closely spaced DBs share a single negative net charge, which can be controllably moved to either DB within the pair. Addition of a nearby negative charge can sufficiently bias or "tip" the potential landscape of the DB pair so as to deterministically place the shared charge to one side of the pair or the other, corresponding to a well-defined binary zero or one. Thus, the pairs become the natural medium to encode binary information by localization of charge, and to perform logic operations. The ability to create and perfectly erase DBs is next used to fabricate and actuate rudimentary circuit elements. We employ the single-electron charge sensitivity of non-contact atomic force microscopy (nc-AFM) to probe the charge configuration and functionality of a fabricated "binary" wire and a logical OR gate.

The nc-AFM images and spectra in Fig. 1a-e characterize the neutral and negative charge states of an isolated DB (corresponding scanning tunneling microscopy (STM) details in *Supporting Figure S1*)[25]. In frequency shift ($\Delta f$) images, recorded at constant height and selected fixed biases (Fig. 1a-c), the background H-Si appear as bright protrusions arranged in the 2×1 surface reconstruction of Si(100)[34] and the DB appears as a variably-dark depression depending on its charge state. The charge transition of the DB is observed as a step-like feature at -350 mV in the blue $\Delta f$ vs. bias voltage spectrum ($\Delta f(V)$, Fig. 1d). Similar steps in $\Delta f(V)$ spectra taken above molecules and atoms are known to correspond to single-electron charge transitions [35–38], as changing the charge state of an entity underneath the tip changes the electrostatic force experienced by the tip, registering as a shift in resonance frequency.

The DB's $\Delta f(V)$ spectrum can be explained qualitatively by considering its (0/-) charge transition level relative to the position of the tip's Fermi level, as the bias voltage is swept. Isolated DBs on a highly n-type doped crystal, as studied here, are negatively charged at zero bias [39–41]. When the tip's Fermi level is energetically above the (0/-) charge transition level of the DB (Fig.

1f), the DB is doubly occupied and therefore negative (region **I** in Fig. 1d). At the step in the *Δf(V)* curve, the tip's Fermi level becomes resonant with the (0/-) charge transition level and the tip extracts an electron from the DB. Because the the (0/-) charge transition level of the DB at this bias voltage lies in the bandgap, the coupling to tip is stronger than to the bulk and there is no efficient re-supply of electrons from the bulk to refill the DB. As a result, the DB is rendered neutral (Fig. 1g). Consequently, the step in *Δf(V)* in Fig. 1d corresponds to the transition of the DB between its negative (right of the step, **I**) and neutral (left of the step, **II**) charge states [25]. The associated difference in *Δf* between the hydrogen atoms (teal curve in Fig. 1d) and the DB at the marked fixed biases explains the contrast differences in the constant height nc-AFM scans of Fig. 1a-c. Profiles taken across the DBs in these images are displayed in Fig. 1e and highlight the difference in *Δf* at the DB's location. The magnitude of the attractive *Δf* shift indicates the charge state of the DB: a negatively charged DB (green and black line, Fig. 1e) appears darker and more attractive than a neutral one (orange line, Fig. 1e).

In Fig. 2, we build upon the established fundamental characteristics of an isolated DB to demonstrate the step-by-step fabrication and characterization of a DB pair and the biasing of that pair by a negative charge positioned nearby. We begin with an isolated DB (Fig. 2a-c). The characterization of the isolated DB qualitatively resembles the observations presented in Fig. 1. In order to show that a second DB is identical to the first, the first DB was erased (controllably capped with an H atom, see *Supporting Information-Creating and Erasing of DBs* for details) thereby allowing the second DB to be studied in isolation. In this way it is ensured that the properties attributed to a DB pair were not due to a second DB of aberrant character. After erasure of the first DB, the new DB was created two lattice sites away (teal marker), and the characterization was repeated (Fig. 2 d-f). The $\Delta f(V)$ spectra of both DBs in isolation (Fig. 2c and f) exhibit a charge transition step at -135 mV, confirming that they are identical.

By recreating the left DB (blue marker, Fig. g,h) a pair is formed. The $\Delta f(V)$ spectra (Fig. 2i) taken at each of the paired dots are identical. When compared with $\Delta f(V)$ spectra obtained above the isolated DBs (Fig. 2c and f), they exhibit a new step at 265 mV. The first step appears at the same energy as that of the isolated DBs (-135 mV). These observations can be understood by noting that below (more negative sample bias) -135 mV both DBs are neutral, between -135 mV and 265 mV only one DB is negatively charged, and above 265 mV both are negatively charged (see *Supporting Information-Pair of two DBs* for more details). Therefore, under the imaging condition of Fig. 2h (0V), only one DB is negatively charged at a time, confirming the pair has one net negative electron. The streaky appearance of the DBs observed in Fig. 2h indicates that the charge switches its position within the pair, an effect detailed elsewhere [42].

Next, a third DB is added five lattice sites away from the teal DB (orange marker, Fig. 2j,k). The $\Delta f$ image now reveals a clear contrast between DBs in the pair (blue and teal marker, Fig. 2k). The $\Delta f(V)$ spectra taken above the three DBs (Fig. 2l) confirm that the blue and orange DBs are both negatively charged at 0 V where the $\Delta f$ image was obtained, while the teal DB remains neutral. As the negatively charged DB appears darker than a neutral one in the $\Delta f$ image, this demonstrates that the charge is biased to reside on the blue DB in the pair. We denote this pair's charge configuration a binary zero and refer to the orange DB as an electrostatic perturber. By subsequently erasing the orange DB and adding a perturber on the left side (five lattice sites away from the blue DB), we demonstrate that the opposite biased case can be achieved (Fig 2m-o). Taken together, these observations show the binary character of the DB pair.

We note here that for both tipped cases, the charge transition of the negative DBs are shifted to -50 mV from -135 mV for an isolated DB (marked by the dashed vertical line). This is due to the two negative DBs being in close enough proximity to interact and mutually shift their charge transition levels to a less negative value. The neutral DB exerts no effect on the negative DBs. In contrast, the presence of two negatively charged DBs next to the neutral DB strongly shifts its charge transition to 395 mV (*Supporting Figure S2f,i*).

The experiments in Fig. 2 can be explained with simple electrostatics and by assuming that negative DBs act as point charges. The red lines in Fig. 2p-r depict the measured energies of the DBs' (0/-) charge transition levels for the isolated, paired, and biased cases, respectively, extracted from the *Δf(V)* spectra of Fig. 2. The tip-induced band bending (see *Supplementary Information-Details on Tip-induced Band Bending*) was calculated and factored in to obtain the corrected DB(0/-) charge transition levels in the absence of the tip (solid blue lines, Fig. 2p-r). The (0/-) charge transition level of the corrected isolated DB (Fig. 2p) is 0.23 eV below the Fermi level. This is in agreement with (0/-) charge transition energies for Si(100)/SiO$_2$ interface dangling bonds (P$_{b0}$ centers) [43] where a value of 0.27±0.1 eV was reported. We note that *ab initio* calculations report significantly lower values for the negative DB state than our obtained value [26,39,44–46].

Figure 2q highlights that closely spaced DB pairs share a single negative charge. As a result of Coulomb repulsion only the (0/-) charge transition level of one of the DBs is below the Fermi level and that of the other DB is lifted above the Fermi-level, rendering it neutral. Adding the perturber DB (Fig. 2r) fixes the charge on one side. From the corrected DB(0/-) charge transition levels we can extract the energy shift as a function of its separation from another negatively charged DB (Fig. 2s). We fit the data with the screened Coulomb energy equation [47],

$$U(r) = \frac{e}{4\pi\varepsilon_0\varepsilon r}e^{-r/L_{TF}},$$

where $e$ is the elementary charge, $\varepsilon$ is effective dielectric constant at the surface, $r$ is the distance between DBs and $L_{TF}$ is the Thomas-Fermi screening length. From the fit (black dashed line, Fig. 2s), $\varepsilon$ and $L_{TF}$ were extracted to be 5.6 and 5 nm, respectively. We note that without factoring in the screening effect, the fit results in a physically invalid negative offset energy (grey dashed line, Fig. 2s).

Figure 2 summarizes the underlying principles of our approach, where we define the two biased configurations of the pair as two binary states. By creating larger ensembles, more complex functionality can be achieved. We now provide two additional examples: a binary wire and a logical OR gate.

Figure 3 demonstrates that the electrostatically determined binary state of a DB pair can be extended over a line formed of many paired DBs. Figure 3a shows an STM image of a wire constructed from eight pairs, with a lone perturber (red marker) on its right. The constant-height $\Delta f$ image below (Fig. 3b) shows that the perturber tips the eight pairs to the left. In Fig. 3c, an additional DB is patterned next to the red DB so that it forms a ninth pair. The $\Delta f$ image (Fig. 3d) demonstrates that such an ensemble, lacking a perturbing electrostatic input becomes self-polarized, with the pairs on either side of its midpoint adopting opposite polarizations (division indicated by the purple dashed line). In Fig. 3e,f a new perturber on the left side of the ensemble, reversing the state. The sequence of images in Fig. 3 demonstrates the basis for a binary wire. Because we are limited at this point to negative charges as inputs, we are restricted to demonstrating the two states of the wire by "pushing" with charges from an input placed at one end of the line or the other. The fixed perturbing charges used in this work are stand-ins for eventual analog wires, with which we anticipate being able to directly electrically control the devices.

Figure 4 shows a logical OR gate can also be achieved through a 2-dimensional arrangement of DB pairs. The gate is composed of three pairs of DBs arranged in a "Y" (Fig. 4a,b). We define the two top-most branches as the gate's inputs, and the lower branch as its output. In the absence of perturbers, the mutual electrostatic repulsion among the electrons within the pairs causes the electrons to localize to the outermost DBs. We define the output state of the gate by only considering the lowermost DB of the output branch (indicated by the red line in the $\Delta f$ images), regardless of the occupation of the other DB. The addition of a perturber below the output branch (Fig. 4d) reverses the output's state and initializes the gate (Fig. 4e-f). In this case we define the neutral state to be 0 and the negatively charged state to be 1, such that the first row of the truth table for the OR gate (Fig. 4c) is established. When a perturber is placed at either of the input branches (Fig. 4g-i, j-l) or both (Fig. 4m-l) the effect of the perturber below the output branch is overcome, enforcing an output of 1 at the designated output DB. Together these configurations satisfy the remaining rows of the truth table.

Continuing efforts will provide a path for large-scale automated patterning and environmental protection through encapsulation [48,49], with variants of the atomic ensembles shown here enabling logical NOT, AND, and fan-out structures. Read-out and connection to outside circuitry can perhaps be accomplished through atomic wire leads [50] connected to SETs [51–53] or quantum point contacts [17,54]. Field controlled computing approaches like ours can, in principle, operate in a very low power and yet ultra-fast regime. Though detailed studies of power requirements and speed of operation remain to be done, it is anticipated that the methods outlined here will be attractive in those regards. Additionally, the room-temperature stability of the DBs, combined with future encapsulation, will allow patterned structures to be taken out of their fabrication environment and transported. Though not demonstrated here, prior work indicates that these structures should also function at room temperature [13,15], warranting future experiments at elevated temperatures.

**Acknowledgments**

We thank Martin Cloutier, Doug Vick, and Mark Salomons for their technical expertise. We thank NRC, NSERC, QSi, Alberta Innovates and Compute Canada for financial support. We thank F. Giessibl for providing us with the tuning forks for building the qPlus sensors. We thank Ken Gordon for valuable suggestions and discussion.

**Author Contributions**

T.H., H.L., M.R., T.D., R.A., W.V. designed and performed the experiments and analyzed the data. T.H., R.A.W., T.D., L.L., W.V., and M.R. co-wrote the paper. L.L and M.R. performed the theoretical modeling. J.L.P. and R.A. contributed to the interpretation and discussion of the results. R.A.W. conceived of and supervised the project. All authors discussed the results and commented on the manuscript.

**Competing Interests**

The authors declare competing financial interests: A patent has been filed on this subject. Some of the authors are affiliated with Quantum Silicon Inc (QSi). QSi is seeking to commercialize atomic silicon quantum dot based technologies.

**Materials and Correspondence**

Correspondence and requests for materials should be addressed to taleana@ualberta.ca; rwolkow@ualberta.ca.

**Figures and Captions**

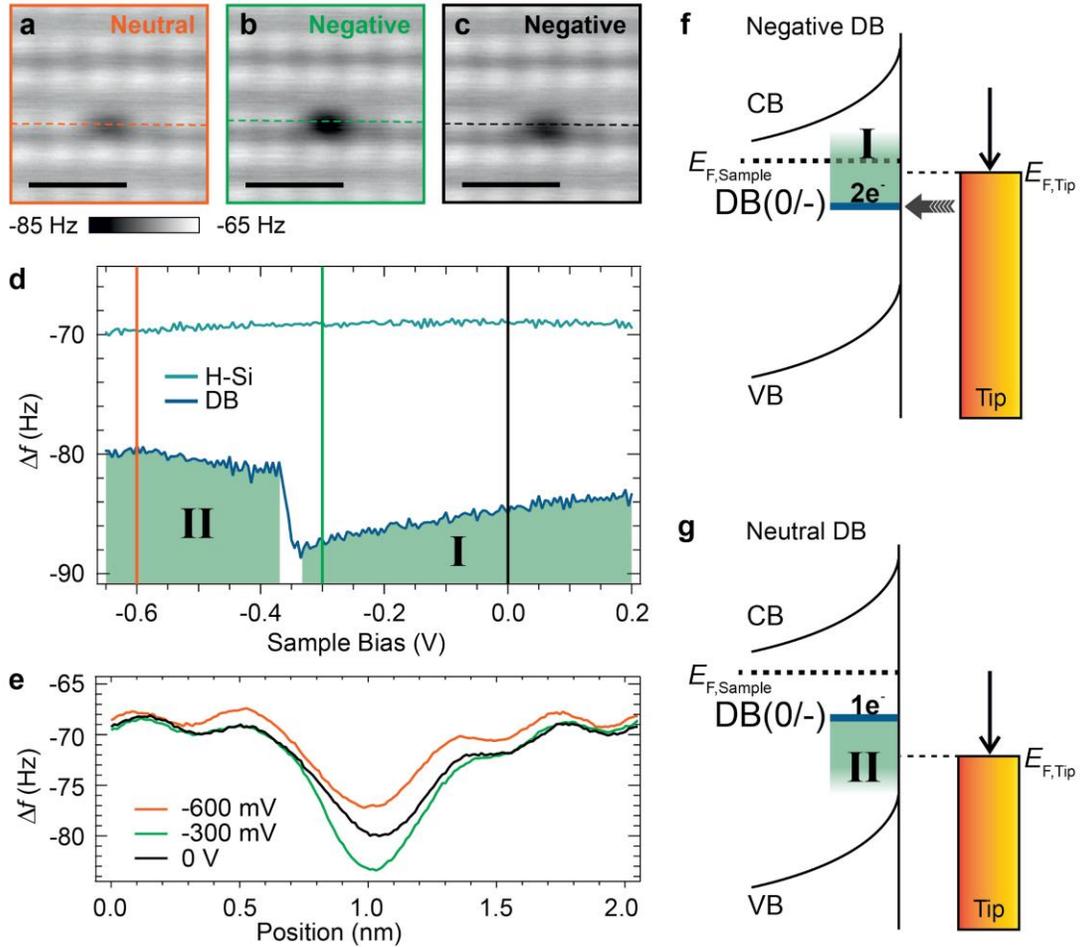

**Figure 1: Probing Charge State Transitions of a Dangling Bond.** (**a-c**) 2×2 nm² constant height $\Delta f$ images of an DB at different bias voltages ($z_{rel}$ = -350 pm, oscillation amplitude = 50 pm, and V = -600 mV (a), V = -300 mV (b), V = 0 V (c)). (**d**) Frequency shift *vs.* sample bias ($\Delta f(V)$) measured above the hydrogen-terminated surface (teal curve) and the DB (blue curve) showing a charge transition step ($z_{rel}$ = -350 pm and osc. amplitude = 50 pm, cf. *Supp. Fig. 1* for STM details). Color-matching vertical lines indicate the fixed sample bias the $\Delta f$ images shown in (a-c) were taken at. (**e**) Scan profiles extracted from (a-c) at the dashed lines as indicated. All scale bars are 1 nm. Green shaded regions I and II denote the negative and neutral charge state bias regions, respectively. (**f**) Qualitative band diagram of the tip-sample system when the DB is negatively charged. The tip Femi level is above the negative to neutral charge transition level DB(0/-) rendering it doubly occupied. (**g**) Band diagram when the DB is neutral showing the tip Fermi level below the DBs charge transition level. Roman numerals in the green shaded regions (f,g) correlate to the bias regions indicated in the DB's $\Delta f(V)$ curve in (d).

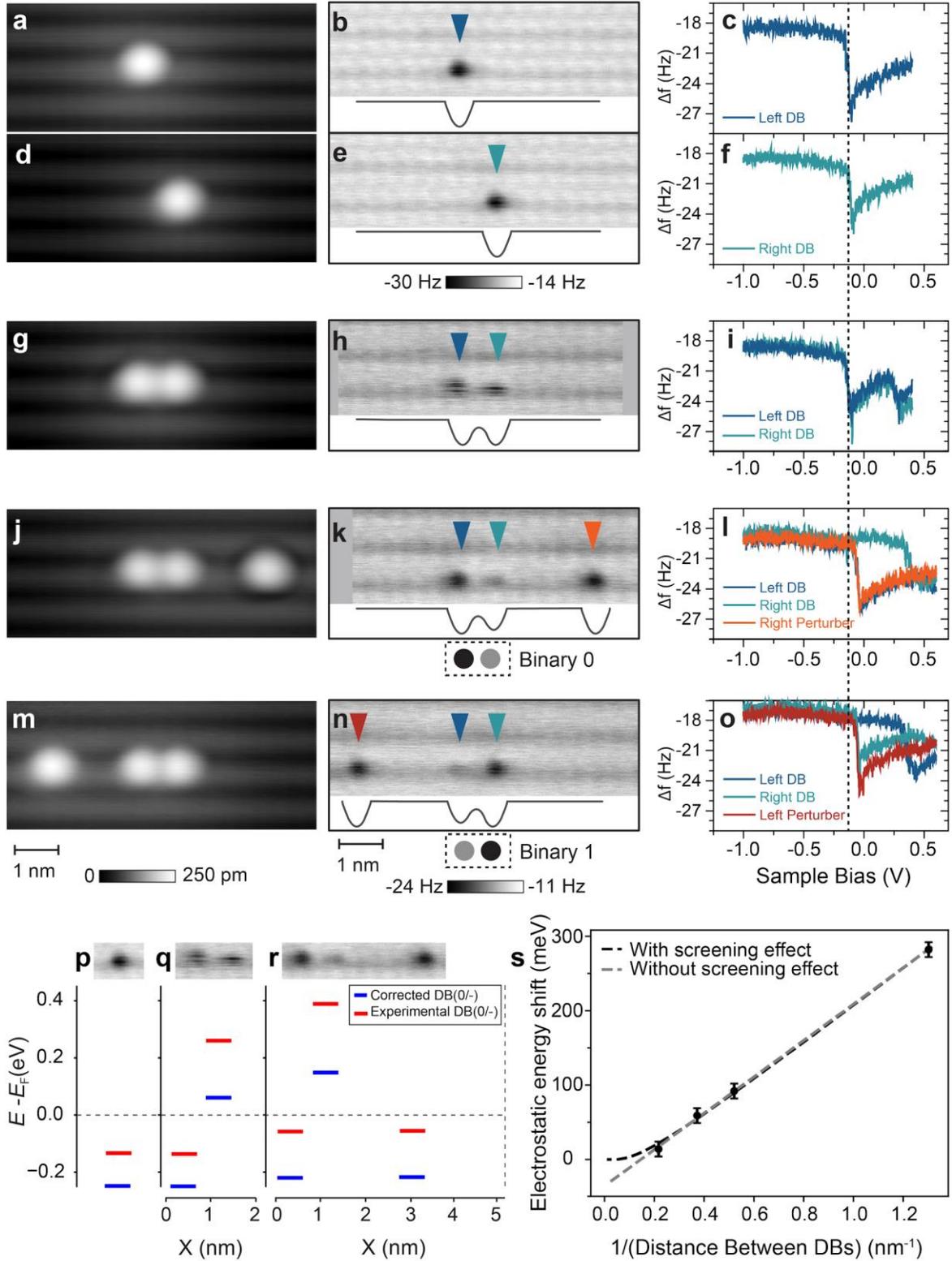

**Figure 2: Biasing of Dangling Bond Structures.** Filled state STM images of the isolated left (**a**), isolated right (**d**), coupled (**g**), biased right (**j**), and biased left (**m**) DB assemblies (V = -1.8 V, I = 50 pA). Corresponding frequency shift (*Δf*) images for each case are shown in (**b**), (**e**), (**h**),

**(k)**, and **(n)**, respectively ($z_{rel}$ = -350 pm for (b), (e) and $z_{rel}$ = -300 pm for (h),(k), and (n), oscillation amplitude = 50 pm, V = 0 V). Qualitative potential energy well sketches are included at the bottom of each *Δf* panel, and the polarized states in (k,n) have their binary representation below. Color coded *Δf(V)* spectra taken on top of the quantum dots in the frequency shift maps are shown in **(c)**, **(f)**, **(i)**, **(l)**, and **(o)** ($z_{rel}$ = -300 pm, oscillation amplitude = 50 pm). The charge transition onset for the isolated DB cases at -135 mV is marked with a vertical dashed line for reference. **(p)**, **(q)**, and **(r)** show the DB(0/-) charge transition levels for the isolated, paired, and perturbed DBs, respectively. Red solid lines are the charge transition level experimentally measured. Blue lines are the corrected energy level in the absence of any tip-induced band bending. **(s)** The electrostatic energy shift of the DB charge transition levels as a function of DB to DB distance for negatively charged DBs. Fits with and without screening factored in are plotted.

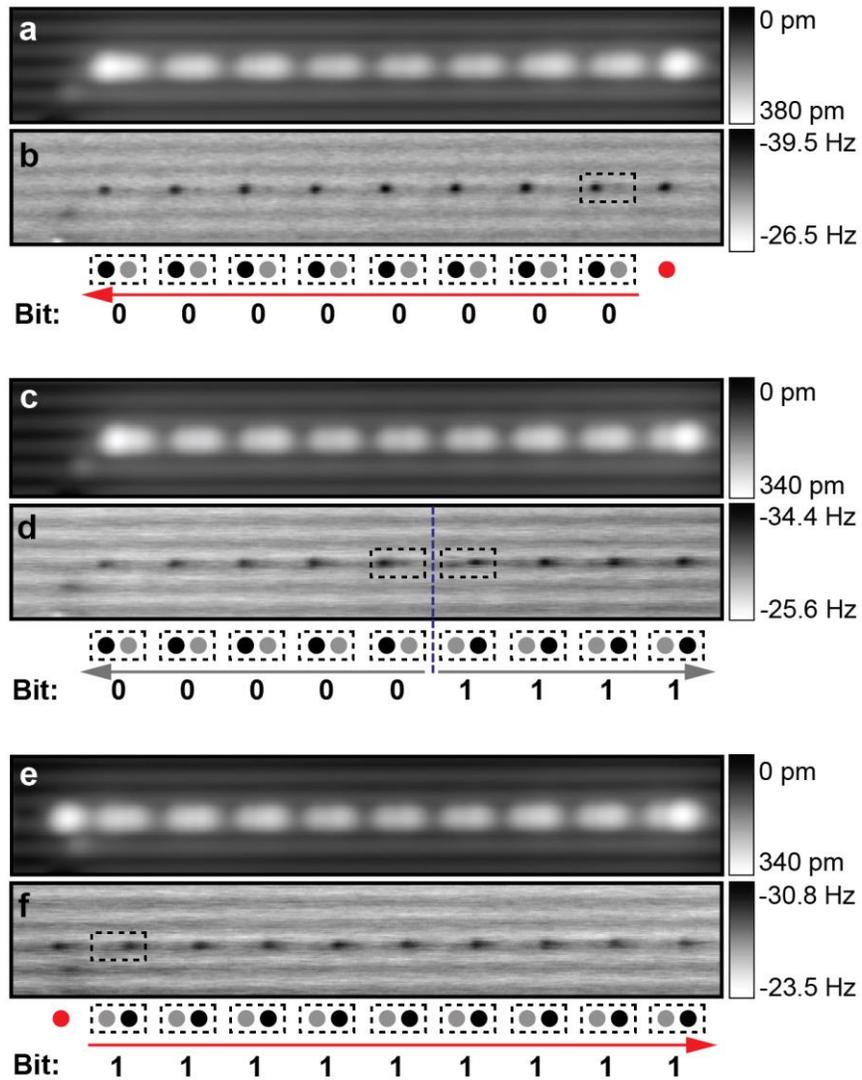

**Figure 3: Information Transmission through a Dangling Bond Binary Wire.** (**a**) Filled states STM image and (**b**) corresponding constant-height *Δf* image of an eight-pair wire with a non-paired perturber DB on the right. (**c**) Symmetric nine-pair wire creating from pairing up the red perturbing DB in (b). (**d**) Constant-height *Δf* image of the nine-pair wire, with symmetry splitting plane marked by a dashed purple line. (**e**) STM image of a nine-pair wire after adding a perturbing DB on the left. (**f**) Constant-height *Δf* image showing the wire binary state under the field of the perturber (red). All STM images were taken at (V = -1.7 V, I = 50 pA). All *Δf* images are 24×3 nm$^2$ in size and were taken at zero bias with a relative tip elevation of $z_{rel}$ = -330 pm and osc. amplitude = 50 pm. Guides are placed below (b), (d), and (f) to show the location and bit state of the pairs.

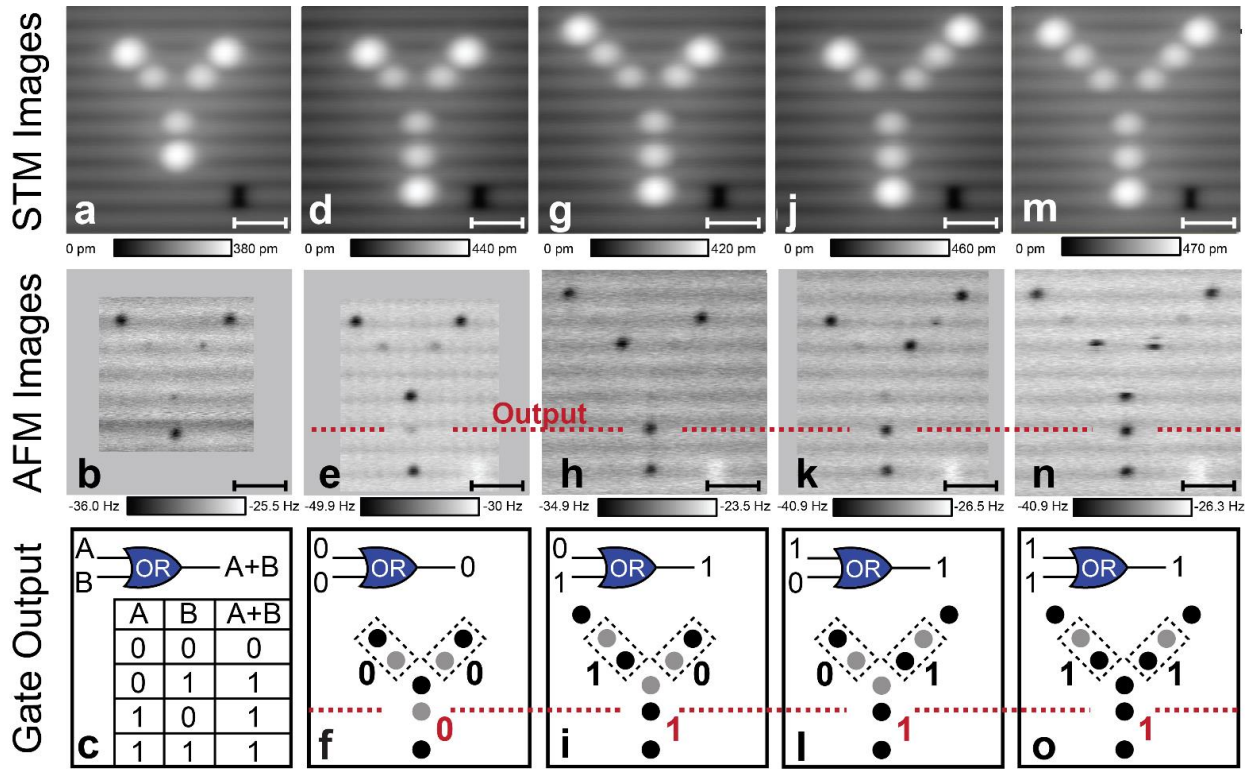

**Figure 4: OR gate Constructed of Dangling Bonds**. **First row:** Constant-current filled state STM images (V = -1.8 V, I = 50 pA) of the OR gate in various actuation states. **Second row:** corresponding constant-height $\Delta f$ images (V = 0 V, $z_{rel}$ = -350 pm) of the gate, displaying electron location as the dark depressions, with the output DB marked in red. **Bottom row:** The complete truth table of an OR gate (c), with models for the four distinct outputs corresponding to the gates displayed vertically above them in rows one and two. Scale bars are 2 nm.

**Supporting Information for**
# Binary Atomic Silicon Logic

Taleana Huff[1,3,*], Hatem Labidi[1,2], Mohammad Rashidi[1], Lucian Livadaru[3], Thomas Dienel[1], Roshan Achal[1,3], Wyatt Vine[1], Jason Pitters[2,3], and Robert A. Wolkow[1,2,3*]

[1]Department of Physics, University of Alberta, Edmonton, Alberta, T6G 2J1, Canada
[2] Nanotechnology Research Centre, National Research Council Canada, Edmonton, Alberta, T6G 2M9, Canada
[3]Quantum Silicon, Inc., Edmonton, Alberta, T6G 2M9, Canada
*Correspondence to: taleana@ualberta.ca rwolkow@ualberta.ca

**Table of contents**
- Measurement System Setup
- Sample Preparation
- Tip Preparation
- Creating and Erasing DBs
- Further details on *Δf(V)* Spectroscopy
    - Pair of Two DBs
    - One DB Perturbing a Pair
    - Symmetrically Perturbed Pair of DBs
- Details on tip-induced band bending

**Measurement System Setup:** Experiments were carried out using a commercial (Scienta-Omicron) qPlus AFM [44] system operating at 4.5 K. Nanonis control electronics and Specs software were used for both STM and AFM data acquisition. For all constant-height frequency shift images and the bias-dependent spectroscopy, $z_{rel} = 0$ p.m corresponds to the relative tip elevation defined by the STM imaging set points on the site of hydrogen-terminated silicon I = 50 pA and V = -1.8 V. The tuning fork had a resonance frequency of 32.8834450 kHz, with a quality factor of 40,000. To minimize drift during AFM image acquisition, the tip was left to settle for 12 hours after approach to allow piezo scanner stabilization. All STM and AFM images are raw data.

**Sample Preparation:** Highly arsenic-doped (~$1.5\times10^{19}$ atom.cm$^{-3}$) Si(100) was used. Sample preparation involved degassing at ~600°C for 12 hours in ultra-high-vacuum (UHV), followed by a series of resistive flash anneals reaching 1250°C to remove oxide, and finally holding the Si substrate at 330°C for two minutes while molecular hydrogen (pressure = $10^{-6}$ Torr) was cracked on a 1600 °C tungsten filament. The series of resistive flash anneals has been shown to reduce surface dopant density, creating a depletion region ~70 nm below the sample surface with a donor concentration ~$10^{18}$ atom.cm$^{-3}$ [41,55].

**Tip Preparation:** A focused ion beam (FIB) was used to cut a micro-tip from electrochemically etched 50 μm polycrystalline tungsten wire, then weld it to the end of a qPlus-style AFM sensor [56]. UHV preparation involved having the oxide layer removed by electron bombardment heating treatments, followed by field evaporation to clean the apex in a field ion microscope (FIM) [57]. Further sharpening was conducted using a FIM nitrogen etching process to obtain the smallest possible tip radius of curvature [57]. Final in-situ tip processing was done through creation of a bare silicon patch through tip induced hydrogen desorption, followed by gentle controlled contacts with the tip on the reactive patch [34].

**Creating and Erasing DBs:** To create an DB, a sharp artifact-free tip is positioned on top of a surface hydrogen atom at 1.3 V and 50 pA, and pulses of 2.0-2.5 V for 10 ms are applied until the hydrogen is removed. Some percentage of the time, the removed single hydrogen atom ends up functionalizing the tip apex. This functionalized tip can be positioned over a DB and mechanically bought towards it to induce a covalent bond and passivate it [31,32].

**Further details on *Δf(V)* Spectroscopy:** Supporting Figure S2 reproduces the bias-dependent frequency shift spectra shown in Fig. 2 of the main text, but with a vertical offset. The vertical offset allows clear discernment of all features and shifts in the graph for the DB pair (Fig. S2a,b), the pair biased to the left (Fig. S2d,e), the pair biased to the right (Fig. S2g,h), and the symmetrically biased pair (Fig. S2j,k). The error to read out the corresponding shifts of the change transition energies are estimated to ±2 mV.

**Pair of two DBs:** For both DBs, the charge transition step is observed at -135 mV, identical to the case of the individual DBs (Fig. 2c,f in the main text). The reason is that only one DB of the pair can harbor an electron in this voltage range (-135 to 265 mV) and the remaining neutral DB does not exert any electric field on its neighboring DB. Furthermore, once the negative charge is localized on one DB, its neighboring DB has its negative to neutral charge transition level instantaneously shifted upward by the Coulomb field of its charged neighbor above the Fermi-level of the sample. This is qualitatively depicted in Supporting Figure S3b. Assuming the blue DB is negative in this example case, as the sample bias is increased further past the second step (≥265 mV), the Fermi level of the tip is raised above the charge transition level DB(0/-) of the teal DB, which captures an electron and becomes negative too (Fig. S3c). This corresponds to both DBs being negative, and to the second step in the *Δf(V)* spectrum at ~+265 mV in Supporting Figure S2c. Both DBs can also be made neutral by reducing the sample bias to bring the tip Fermi level below the charge transition levels for both DBs, as shown in Fig. S3a.

**One DB Perturbing a Pair :** We next consider the "2+1" experiment with a single negative DB perturbing a pair as shown in Figs. 2j-o of the main text. The *Δf(V)* spectra with added offset are reproduced in Supporting Figs. S2b,c. First, we examine the curves for the perturber and the paired DB furthest from the perturber (orange and dark blue in S2b, red and teal in S2c). In all cases the sharp charge transition step is observed at approximately -50 mV. Contrasting this with the charge transition value for a lone DB (-135 mV), it is apparent that an absolute shift of 85 mV occurred. This shift can be explained by the presence of the negative charge at the perturbing DB (for bias values between -50 to 395 mV), electrostatically shifting all nearby DB levels. In other words, referencing specifically S2b, when a $\Delta f(V)$ spectrum is taken with the tip over the blue DB, the step is at -50 mV because the far orange perturbing DB is negative and shifts the level 85 mV closer to zero. The related qualitative band-bending diagram is depicted in Fig. S4b (corresponding to region **II** in the reproduced $\Delta f(V)$ spectra at the bottom). When both orange and blue are negative (-50 mV to 395 mV), they cause a potential energy increase for the middle teal DB pushing it's DB (0/-) level above the sample Fermi level, leaving it neutral. Hence, when taking a $\Delta f(V)$ spectrum over the teal DB, the tip level must be swept to ~+395 mV before it can cross the teal's DB (0/-) transition level and capture an additional electron to become negative (Fig. S4c). As for the pair, the tip level can also be swept below the charge transition levels for all three ($\leq$ -50 mV) rendering them all neutral (Fig. S4a).

**Symmetrically Perturbed Pair of DBs:** We finally consider the "1+2+1" experiment where the two close DBs were symmetrically perturbed by two DBs. The "outside" DBs (red and orange labels, Supporting Figure S2j-l) show a relatively small shift of 22 mV of (0/-) charge transition levels, due to the distance of 12 lattice sites between the perturbers. Furthermore, the presence of two negative charges raises the charge transition levels of the inner dots (blue and teal) to +80 mV.

**Details on Tip-induced Band Bending:** During AFM imaging, even at zero applied bias, the tip affects the electrostatic potential and, under certain conditions, the occupation of the DB system. The effect can be linked to the contact potential difference, whose source is the difference between the work functions of the tip and the sample. For a tungsten tip, the work function varies with the crystal orientation and is typically taken to be between 4.5 and 5 eV. For a silicon sample, the work function varies significantly with doping type and level. For n-type Si at low temperature, the work function is close to the electron affinity, typically 4.05 eV, being less than this value for a degenerate sample, and greater for a non-degenerate sample. In our case, for our tungsten tip the work function was assumed to be 4.5 eV, while for the silicon sample, taking into account the low-temperature and a surface dopant concentration of $10^{18}$ cm$^{-3}$, it was estimated at 4.1 eV.

This difference in work functions leads to band-bending locally under the tip apex that can shift the electronic levels of dangling bonds, thereby potentially emptying or filling them. We refer to the shift as tip induced band bending (TIBB). TIBB is strongest immediately under the tip apex. For the above quoted work functions, the TIBB is in the upward direction at zero applied bias voltage, *i.e.*, levels get shifted upward with respect to the sample fermi level $E_{F,sample}$. While the contact potential difference is a constant, the TIBB changes with both tip-sample separation, as well as applied tip-sample bias (See Supporting Figure S5). If an electronic level for a DB is shifted above $E_{F,sample}$, then it cannot stay filled (occupied) in electrochemical equilibrium.

The exact value of TIBB depends not only on the contact potential $V_c$, but also on the following parameters: sample doping level ($N_d$), tip-sample bias ($V_{st}$), tip shape, apex radius ($R_t$), and the distance (height) between the tip and the surface ($d$). The TIBB in Supporting Figure S5 was calculated using our best estimates for the above parameters. A 3D finite-element Poisson equation solver was employed to calculate the TIBB using methodology described in reference [58].

**Supporting Figures**

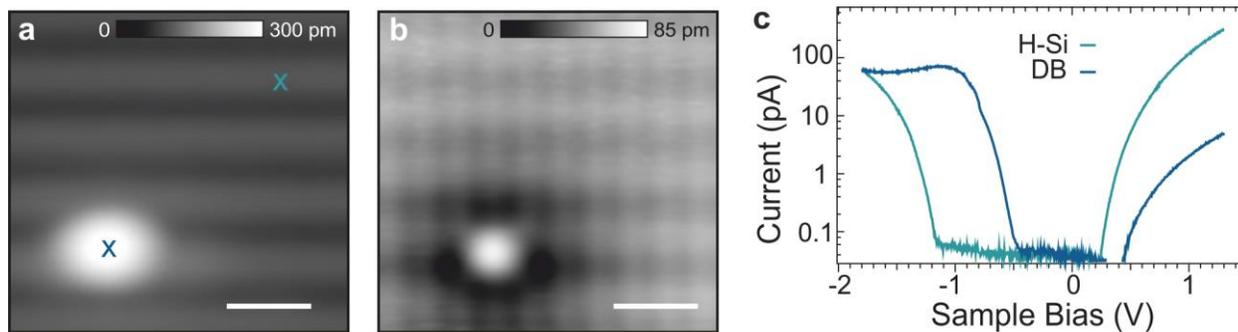

**Supporting Figure S1: STM characterization of a Dangling Bond.** (**a**) 4×4 nm² filled states STM image (V = -1.8 V, I = 50 pA) and (**b**) 4×4 nm² empty states STM image (V = 1.3 V, I = 50 pA) of a DB. (**c**) Tunneling current *vs.* sample bias (*I(V)*) spectroscopy plotted in log scale of the DB (blue curve) and hydrogen-terminated surface (teal curve). Spectroscopy positions indicated in (a).

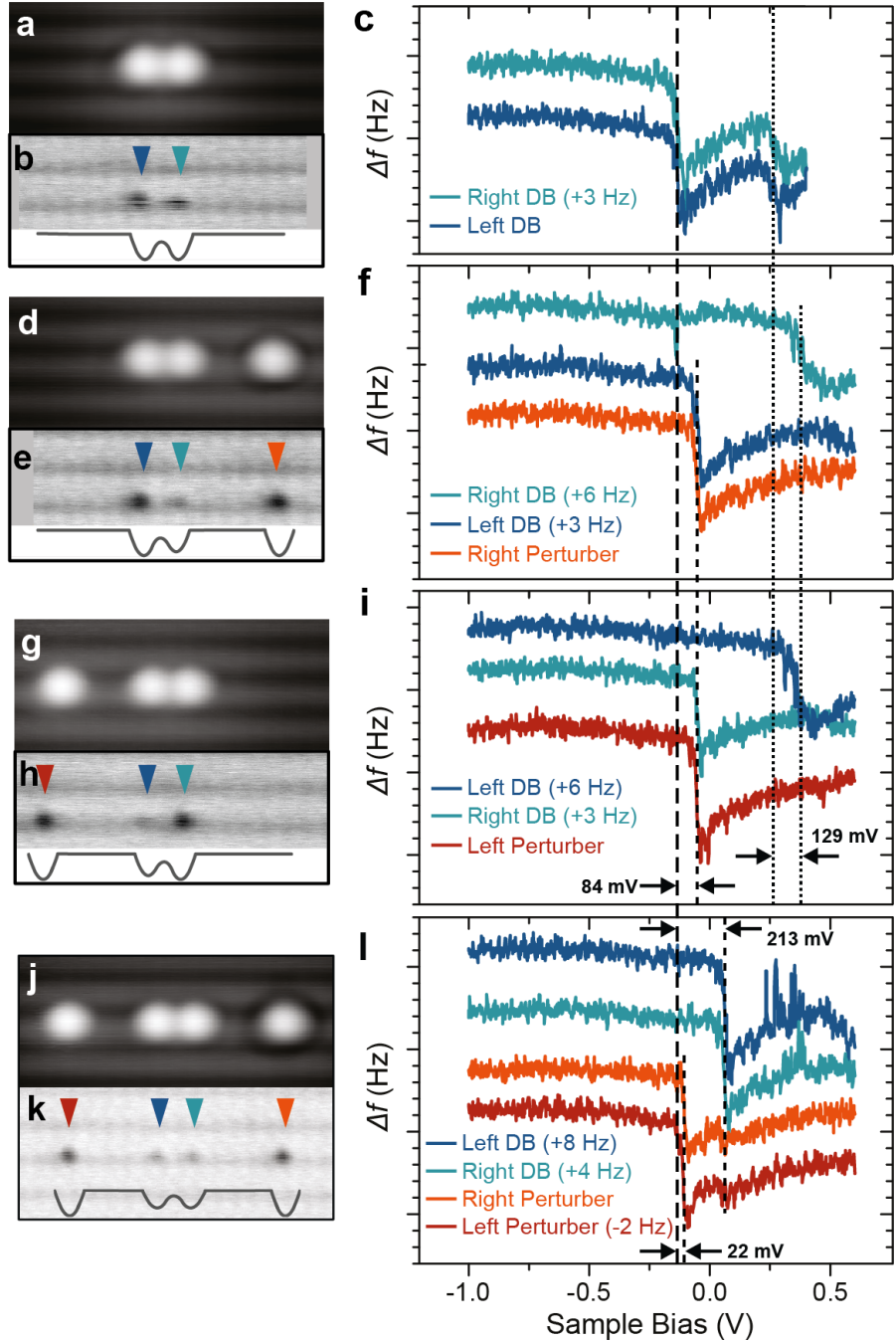

**Supporting Figure S2: Frequency Shift Spectroscopy in Dangling Bond Structures.** Color coded spectra from main text Figure 2 reproduced with vertical offsets for the *Δf(V)* to show key features for the pair **(a-c)**, left tipped **(d-f)**, right tipped **(g-i)**, and symmetric **(j-l)** cases (being STM, constant-height AFM, and vertically offset *Δf(V)*, respectively for each case). The charge transition onset for the isolated DB cases, taken from the pair in (c), is marked with a vertical long-dashed line for reference. A short-dashed line, only in (f) and (i), indicates the shifted charge transition in the presence of one additional charge (the perturber). The finely dotted lines indicate the charge transition onset for bringing in the second charge to the pair (c) as well as for the perturbed dot (f,i) in the presence of the charge of the perturber. In (l), the shifted charge transition onset of the perturbers in the presence of its symmetric perturbing partner is marked by a short-dashed line only running over the orange and red spectra. The transition for bringing in an additional electron for the middle pair is marked by the short-dashed line. STM images in (a),(d),(g), and (j) were taken with V = -1.8 V, I = 50 pA. The $\varDelta f$ images in (b), (e), (h) and (k) were taken with $z_{rel}$ = -300 pm, oscillation amplitude = 50 pm, and V=0 V. All $\varDelta f(V)$, were also taken at $z_{rel}$ = -300 pm and oscillation amplitude = 50 pm (the same as reported in main text Figure 2).

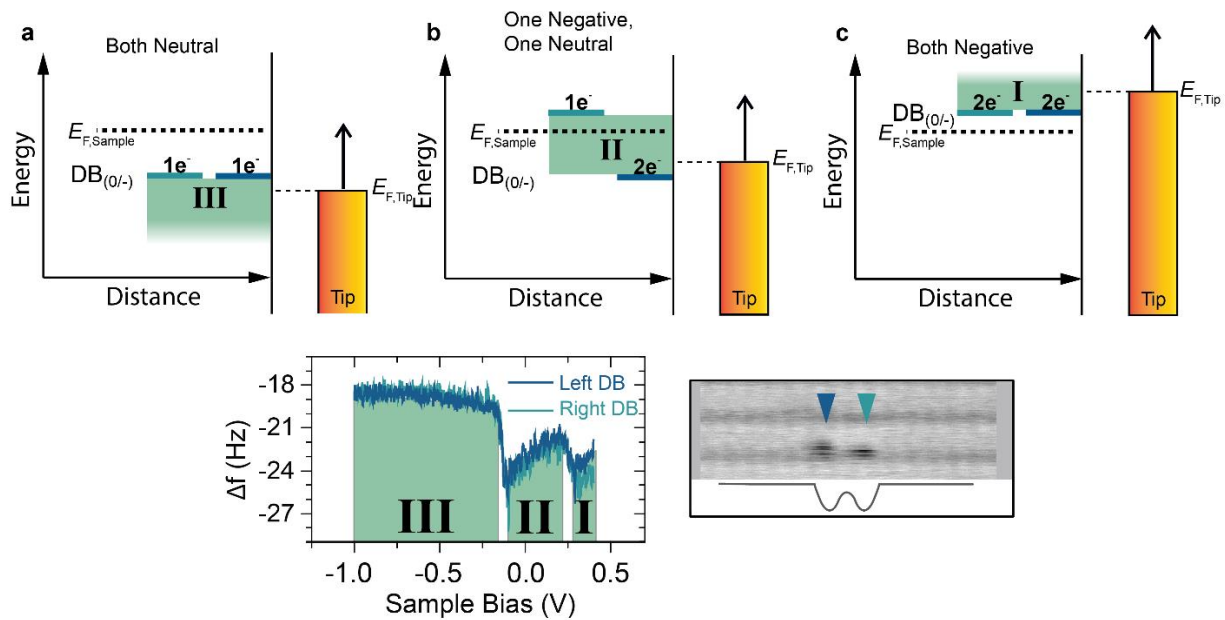

**Supporting Figure S3: Diagrams for Charge Transitions in a Dangling Bond Pair.**
**(a)** Diagram of the system when both DBs are neutrally charged. The tip is assumed to be positioned above the dark blue DB. The DB's negative to neutral charge transition levels are plotted on the left, and are color coded to the $\Delta f(V)$ reproduced from main text Figure 2 below. The Fermi level for tip and sample are given by the dotted lines. The tip Fermi level is below both charge transition levels, meaning both are singly occupied. This corresponds to region III for sample bias ≤ less than -116 mV. **(b)** Diagram for the same system when the sample bias is between -116mV to 260mV. Only the blue DB is negative. The teal DB is neutral, as its charge transition level has been shifted above the Fermi level of the sample from the negative charge of the blue DB. This corresponds to region II in the $\Delta f(V)$ spectra. **(c)** Diagram for sample bias values greater than 260 mV (region I) in the $\Delta f(V)$ spectrum of two closely spaced DBs where both are negatively charged. The Fermi level of the tip is now above the negative to neutral charge transition level DB(0/-) of both DBs, rendering them both negative.

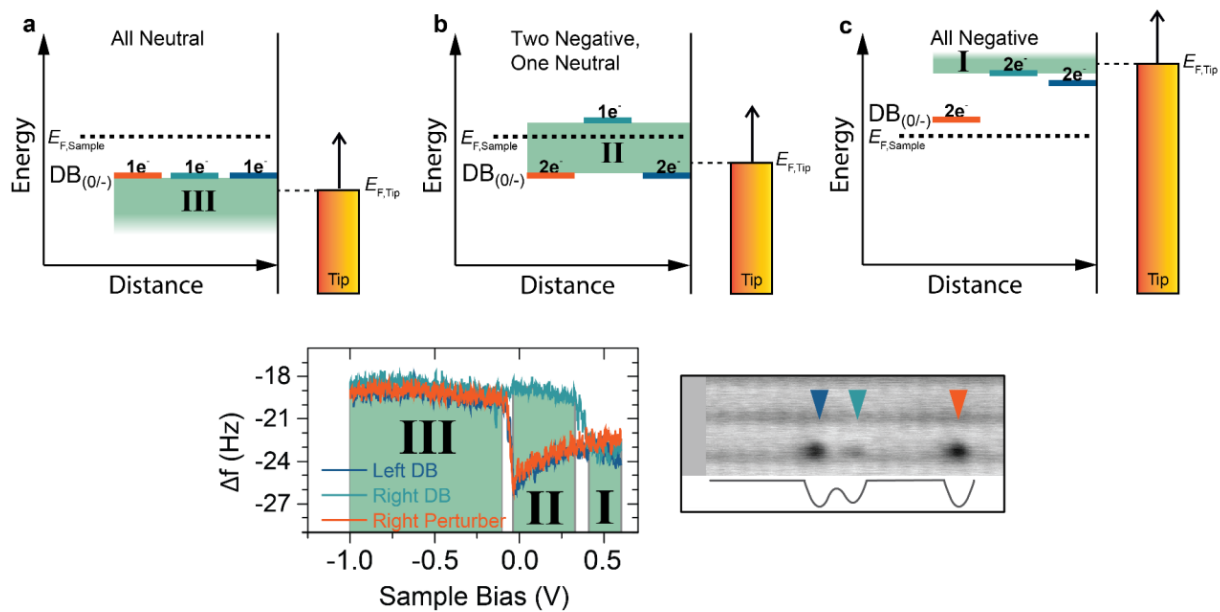

**Supporting Figure S4: Diagrams for Charge Transitions in a Biased Pair of Dangling Bonds. (a)** Diagram of the system when all DBs are neutrally charged. The DB's negative to neutral charge transition levels DB(0/-) are color coded to the $\Delta f(V)$ spectrum at the bottom reproduced from main text Fig. 2. The Fermi level for tip and sample are given by the dotted lines. The tip Fermi level is below all charge transition levels, meaning all are singly occupied. This corresponds to region III for bias ≤ -60 mV. **(b)** The diagrams for the same system when the sample bias is decreased to between -60 mV to 377 mV. The perturbing orange DB and blue DB are both negative, lifting the level for the teal above the Fermi level of the sample and rendering it neutral. This corresponds to region II in the $\Delta f(V)$ spectra. **(c)** Diagrams for sample bias ≥ 377mV (region I) in. The Fermi level of the tip is now above charge transition level of all DBs, rendering them all negative.

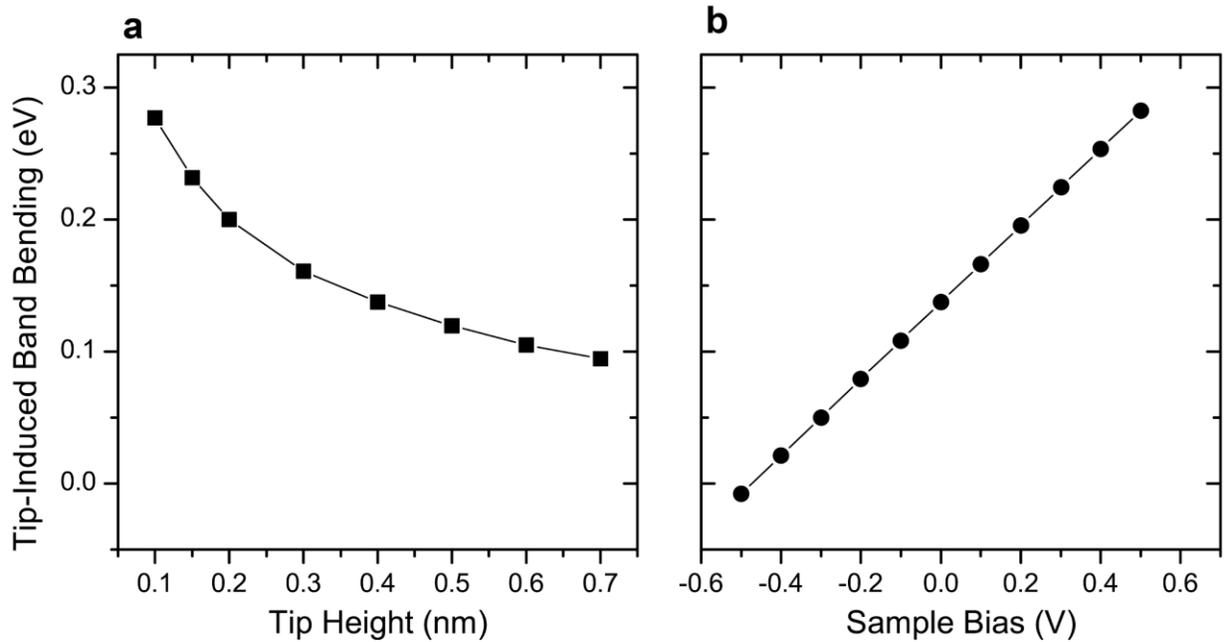

**Supporting Figure S5: Calculated Tip-Induced Band Bending as a Function of Height and Bias.** (**a**) Tip-induced band bending as a function of tip-sample height. No bias is applied between tip and sample. (**b**) Tip-induced band bending as a function of sample bias for a fixed tip-sample separation of 0.4 nm. For both plots, we assumed a donor concentration of $10^{18}$ cm$^{-3}$ at the surface, gradually increasing to $2\times10^{19}$ cm$^{-3}$ in the bulk over a range of approximately 100 nm, a work function difference between tip and sample of 0.4 eV, and a tip radius of 10 nm.

**References**

1. Heinrich, A. J., Lutz, C. P., Gupta, J. A. & Eigler, D. M. Molecule Cascades. *Science* **298,** 1381–1387 (2002).
2. de Silva, A. P., Uchiyama, S., Vance, T. P. & Wannalerse, B. A supramolecular chemistry basis for molecular logic and computation. *Coord. Chem. Rev.* **251,** 1623–1632 (2007).
3. Soe, W.-H. *et al.* Demonstration of a NOR logic gate using a single molecule and two surface gold atoms to encode the logical input. *Phys. Rev. B* **83,** 155443 (2011).
4. Collier, C. P. *et al.* Electronically Configurable Molecular-Based Logic Gates. *Science* **285,** 391–394 (1999).
5. de Silva, A. P. & Uchiyama, S. Molecular logic and computing. *Nat. Nanotechnol.* **2,** 399 (2007).
6. Wang, Y. *et al.* Field-effect transistors made from solution-grown two-dimensional tellurene. *Nat. Electron.* **1,** 228–236 (2018).
7. Joachim, C., Gimzewski, J. K. & Aviram, A. Electronics using hybrid-molecular and mono-molecular devices. *Nature* **408,** 541 (2000).
8. Khajetoorians, A. A., Wiebe, J., Chilian, B. & Wiesendanger, R. Realizing All-Spin–Based Logic Operations Atom by Atom. *Science* **332,** 1062–1064 (2011).
9. Fresch, B., Bocquel, J., Rogge, S., Levine, R. D. & Remacle, F. A Probabilistic Finite State Logic Machine Realized Experimentally on a Single Dopant Atom. *Nano Lett.* **17,** 1846–1852 (2017).
10. Amlani, I. *et al.* Digital Logic Gate Using Quantum-Dot Cellular Automata. *Science* **284,** 289–291 (1999).
11. Imre, A. *et al.* Majority Logic Gate for Magnetic Quantum-Dot Cellular Automata. *Science* **311,** 205–208 (2006).
12. Lent, C. S. & Tougaw, P. D. A device architecture for computing with quantum dots. *Proc. IEEE* **85,** 541–557 (1997).
13. Wolkow, R. A. *et al.* Silicon Atomic Quantum Dots Enable Beyond-CMOS Electronics. in *Field-Coupled Nanocomputing: Paradigms, Progress, and Perspectives* Pg. 33–58 (Springer Berlin Heidelberg, 2014)
14. Orlov, A. O., Amlani, I., Bernstein, G. H., Lent, C. S. & Snider, G. L. Realization of a Functional Cell for Quantum-Dot Cellular Automata. *Science* **277,** 928–930 (1997).
15. Haider, M. B. *et al.* Controlled Coupling and Occupation of Silicon Atomic Quantum Dots at Room Temperature. *Phys. Rev. Lett.* **102,** 046805 (2009).
16. Gorman, J., Hasko, D. G. & Williams, D. A. Charge-Qubit Operation of an Isolated Double Quantum Dot. *Phys. Rev. Lett.* **95,** 090502 (2005).
17. Kim, D. *et al.* Quantum control and process tomography of a semiconductor quantum dot hybrid qubit. *Nature* **511,** 70 (2014).
18. Schedelbeck, G., Wegscheider, W., Bichler, M. & Abstreiter, G. Coupled Quantum Dots Fabricated by Cleaved Edge Overgrowth: From Artificial Atoms to Molecules. *Science* **278,** 1792–1795 (1997).
19. Bayer, M. *et al.* Coupling and Entangling of Quantum States in Quantum Dot Molecules. *Science* **291,** 451–453 (2001).
20. Lent, C. S., Tougaw, P. D., Porod, W. & Bernstein, G. H. Quantum cellular automata.

*Nanotechnology* **4,** 49 (1993).
21. Landauer, R. Minimal Energy Requirements in Communication. *Science* **272,** 1914–1918 (1996).
22. Mathur, N. Beyond the silicon roadmap. *Nature* **419,** 573 (2002).
23. Shibata, K., Yuan, H., Iwasa, Y. & Hirakawa, K. Large modulation of zero-dimensional electronic states in quantum dots by electric-double-layer gating. *Nat. Commun.* **4,** 2664 (2013).
24. Taucer, M. *et al.* Single-Electron Dynamics of an Atomic Silicon Quantum Dot on the H-Si(100) - (2x1) Surface. *Phys. Rev. Lett.* **112,** 256801 (2014).
25. Rashidi, M. *et al.* Resolving and Tuning Carrier Capture Rates at a Single Silicon Atom Gap State. *ACS Nano* **11,** 11732–11738 (2017).
26. Scherpelz, P. & Galli, G. Optimizing surface defects for atomic-scale electronics: Si dangling bonds. *Phys. Rev. Mater.* **1,** 021602 (2017).
27. Schwalb, C. H., Dürr, M. & Höfer, U. High-temperature investigation of intradimer diffusion of hydrogen on Si(001). *Phys. Rev. B* **82,** 193412 (2010).
28. McEllistrem, M., Allgeier, M. & Boland, J. J. Dangling Bond Dynamics on the Silicon (100)-2x1 Surface: Dissociation, Diffusion, and Recombination. *Science* **279,** 545–548 (1998).
29. Achal, R. *et al.* Lithography for Robust and Editable Atomic-scale Silicon Devices and Memories (In Press, Details to be Provided). *Nat. Commun.* (2018).
30. Lyding, J. W., Shen, T.-C., Abeln, G. C., Wang, C. & Tucker, J. R. Nanoscale patterning and selective chemistry of silicon surfaces by ultrahigh-vacuum scanning tunneling microscopy. *Nanotechnology* **7,** 128 (1996).
31. Huff, T. R. *et al.* Atomic White-Out: Enabling Atomic Circuitry through Mechanically Induced Bonding of Single Hydrogen Atoms to a Silicon Surface. *ACS Nano* **11,** 8636–8642 (2017).
32. Pavliček, N., Majzik, Z., Meyer, G. & Gross, L. Tip-induced passivation of dangling bonds on hydrogenated Si(100)-2×1. *Appl. Phys. Lett.* **111,** 053104 (2017).
33. Rashidi, M. & Wolkow, R. A. Autonomous Scanning Probe Microscopy in Situ Tip Conditioning through Machine Learning. *ACS Nano* (2018). doi:10.1021/acsnano.8b02208
34. Labidi, H. *et al.* Indications of Chemical Bond Contrast in AFM Images of a Hydrogen-Terminated Silicon Surface. *Nat. Commun.* **8,** 14222 (2017).
35. Bussmann, E. & Williams, C. C. Single-Electron Tunneling Force Spectroscopy of an Individual Electronic State in a Nonconducting Surface. *Appl. Phys. Lett.* **88,** 263108 (2006).
36. Steurer, W. *et al.* Manipulation of the Charge State of Single Au Atoms on Insulating Multilayer Films. *Phys. Rev. Lett.* **114,** 036801 (2015).
37. Stomp, R. *et al.* Detection of Single-Electron Charging in an Individual InAs Quantum Dot by Noncontact Atomic-Force Microscopy. *Phys. Rev. Lett.* **94,** 056802 (2005).
38. Wagner, C. *et al.* Scanning Quantum Dot Microscopy. *Phys. Rev. Lett.* **115,** 026101 (2015).
39. Schofield, S. R. *et al.* Quantum Engineering at the Silicon Surface Using Dangling Bonds. *Nat. Commun.* **4,** 1649 (2013).
40. Livadaru, L., Pitters, J., Taucer, M. & Wolkow, R. A. Theory of nonequilibrium single-electron dynamics in STM imaging of dangling bonds on a hydrogenated silicon surface.

*Phys. Rev. B* **84,** 205416 (2011).
41. Labidi, H. *et al.* Scanning Tunneling Spectroscopy Reveals a Silicon Dangling Bond Charge State Transition. *New J. Phys.* **17,** 073023 (2015).
42. Rashidi, M. *et al.* Initiating and monitoring the evolution of single electrons within atom-defined structures. Preprint at https//arxiv.org/abs/1709.10091 (2017).
43. Gerardi, G. J., Poindexter, E. H., Caplan, P. J. & Johnson, N. M. Interface traps and Pb centers in oxidized (100) silicon wafers. *Appl. Phys. Lett.* **49,** 348–350 (1986).
44. Blomquist, T. & Kirczenow, G. Controlling the charge of a specific surface atom by the addition of a non-site-specific single impurity in a Si nanocrystal. *Nano Lett.* **6,** 61–65 (2006).
45. Livadaru, L. *et al.* Dangling-bond charge qubit on a silicon surface. *New J. Phys.* **12,** 83018 (2010).
46. Bellec, A. *et al.* Electronic properties of the n-doped hydrogenated silicon (100) surface and dehydrogenated structures at 5 K. *Phys. Rev. B* **80,** 245434 (2009).
47. Schubert, E. F. *Doping in III-V Semiconductors*. Ch. 1. (Cambridge Univ. Press 2010).
48. Bunch, J. S. *et al.* Impermeable Atomic Membranes from Graphene Sheets. *Nano Lett.* **8,** 2458–2462 (2008).
49. Sordes, D. *et al.* Nanopackaging of Si(100)H Wafer for Atomic-Scale Investigations. in *On-Surface Atomic Wires and Logic Gates : Updated in 2016 Proceedings of the International Workshop on Atomic Wires, Krakow, September* Pg. 25–51 (Springer International Publishing, 2017)
50. Engelund, M. *et al.* Search for a Metallic Dangling-Bond Wire on n-Doped H-Passivated Semiconductor Surfaces. *J. Phys. Chem. C* **120,** 20303–20309 (2016).
51. Barthel, C. *et al.* Fast sensing of double-dot charge arrangement and spin state with a radio-frequency sensor quantum dot. *Phys. Rev. B* **81,** 161308 (2010).
52. Fuechsle, M. *et al.* A single-atom transistor. *Nat. Nanotechnol.* **7,** 242–246 (2012).
53. Prager, A. A., Orlov, A. O. & Snider, G. L. Integration of CMOS, single electron transistors, and quantumdot cellular automata. in *2009 IEEE Nanotechnology Materials and Devices Conference* 54–58 (2009). doi:10.1109/NMDC.2009.5167548
54. Goan, H.-S., Milburn, G. J., Wiseman, H. M. & Bi Sun, H. Continuous quantum measurement of two coupled quantum dots using a point contact: A quantum trajectory approach. *Phys. Rev. B* **63,** 125326 (2001).
55. Pitters, J. L., Piva, P. G. & Wolkow, R. A. Dopant depletion in the near surface region of thermally prepared silicon (100) in UHV. *J. Vac. Sci. Technol. B, Nanotechnol. Microelectron. Mater. Process. Meas. Phenom.* **30,** 21806 (2012).
56. Labidi, H. *et al.* New fabrication technique for highly sensitive qPlus sensor with well-defined spring constan. *Ultramicroscopy* **158,** 33–37 (2015).
57. Rezeq, M., Pitters, J. & Wolkow, R. Tungsten Nanotip Fabrication by Spatially Controlled Field-Assisted Reaction with Nitrogen. *J. Chem. Phys.* **124,** 204716 (2006).
58. Ryan, P. M., Livadaru, L., DiLabio, G. A. & Wolkow, R. A. Organic Nanostructures on Hydrogen-Terminated Silicon Report on Electric Field Modulation of Dangling Bond Charge State. *J. Am. Chem. Soc.* **134,** 12054–12063 (2012).